\documentclass[twocolumn,amsmath,amssymb,showpacs]{revtex4}\newcommand{\JHEPonly}[1]{}\newcommand{\PRDonly}[1]{#1}

\PRDonly{\usepackage[dvips]{graphicx}}

\def\imo{i}

\begin{document}

\PRDonly{
\title{Passage of radiation through wormholes of arbitrary shape}
\author{R. A. Konoplya}\email{konoplya@tap.scphys.kyoto-u.ac.jp}
\affiliation{Department of Physics, Kyoto University, Kyoto 606-8501, Japan  \\
\& \\
Theoretical Astrophysics, Eberhard-Karls University of T\"{u}bingen, T\"{u}bingen 72076, Germany}
\author{A. Zhidenko}\email{zhidenko@fma.if.usp.br}
\affiliation{Instituto de F\'{\i}sica, Universidade de S\~{a}o Paulo \\
C.P. 66318, 05315-970, S\~{a}o Paulo-SP, Brazil}

\begin{abstract}
We study quasinormal modes and scattering properties via calculation of $S$-matrix for scalar and electromagnetic fields propagating in the background of spherically symmetric and axially symmetric traversable Lorentzian wormholes of a generic shape. Such wormholes are described by the general Morris-Thorne ansatz. The properties of quasinormal ringing and scattering are shown to be determined by the behavior of the wormhole's shape function $b(r)$ and shift-factor $\Phi(r)$ \emph{near the throat}. In particular, wormholes with the shape function $b(r)$ such that $b'(r) \approx 1$, have very long-lived quasinormal modes in the spectrum. We have proved that the axially symmetric traversable Lorentzian wormholes, unlike black holes and other compact rotating objects, does not allows for super-radiance. As a by product we have shown that the 6-th order WKB formula used for scattering problems of black or wormholes gives quite high accuracy and thus can be used for quite accurate calculations of the Hawking radiation processes around various black holes.
\end{abstract}

}
\maketitle

\section{Introduction}

A wormhole is usually a kind of ``shortcut'' through space-time connecting distant regions of the same universe or two universes. General Relativity though does not forbid wormholes, requires presence of some exotic matter for their existence. A wide class of wormholes respecting the strong equivalence principle is called \emph{Lorentzian wormholes}, whose space-time, unlike that of \emph{Euclidean wormholes}, is locally a Lorentzian manifold. A physically interesting subclass of Lorentzian wormholes consists of \emph{traversable wormholes} through which a body can pass for a finite time. This last class of wormholes will be probed here by the passage of test fields in the wormhole vicinity and observing their response in the form of the waves scattering and quasinormal ringing. However, before discussing the scattering processes, let us briefly review some basic features and observational opportunities of wormholes.

The static spherically symmetric Lorentzian traversable wormholes can be modeled by a generic Morris-Thorne metric \cite{Morris-Thorne}, which is fully determined by the two functions: the shape function $b(r)$ and the redshift function $\Phi(r)$. The axially symmetric traversable wormholes can be described by generalization of the Morris-Thorne ansatz, suggested by Teo \cite{Teo:1998dp}. In order to support such a wormhole gravitationally stable against small perturbations of the space-time one needs an exotic kind of matter which would violate the weak energy condition. The latter requires that for every future-pointing timelike vector field, the matter density observed by the corresponding observers is always non-negative. Thus for supporting a traversable wormhole one needs a source of matter with negative density such as, for example, \emph{dark energy} \cite{Ida:1999an}. Wormholes might appear as a result of inflation of primordial microscopic wormholes to a macroscopic size during the inflationary phase of the Universe. If wormholes exist, they could appear as a bubble through which one could see new stars. Massive, rotating wormholes surrounded by some luminous matter could have a visible ``accretion disk'' near them, similar to that surrounding supermassive black holes. Thus, gravitational lensing should give some opportunities for detecting wormholes \cite{Nandi:2006ds,Safonova:2001vz}. General features of various wormholes were investigated in a great number of papers and we would like to refer the reader to only a recent few, where further references can be found \cite{Balakin:2010ar,Richarte:2010bd,Bandyopadhyay:2009ek,Bronnikov:2010hu,Usmani:2010cd,Sarbach:2010we,Mimoso:2010yp,Hayward:2002sz,Lobo:2009du}.

As we mentioned before, stability of wormholes is the crucial criterium for their existence \cite{Gonzalez:2008wd,Gonzalez:2008xk,Doroshkevich:2008xm,Amendo}. In order to judge about a wormholes stability, one has to treat the complete set of the Einstein equations with some non vanishing energy momentum tensor representing the matter source. Thereby, stability depends on particular geometry of the wormhole metric as well as on the equation of state for matter. In this paper, we shall set up a more generic problem: implying the existence of some matter that provides stability of generic traversable wormholes, whose metric is given by some generic redshift $\Phi(r)$ and shape $b(r)$ functions; we are interested in scattering processes of test fields in the vicinity of such wormholes. In other words, we shall consider the passage of test radiation, which does not influence the wormhole geometry, through wormholes of various shapes. Thus, our aim is to learn how the wormhole shape correlates with the scattering properties of various fields in its background. This is a kind of probing of a wormhole's geometry by test fields.

The scattering phenomenon around wormholes can be descried by a number of characteristics, two of which, the \emph{quasinormal modes} and the \emph{\emph{S-matrix}}, we shall consider here. The quasinormal modes are proper oscillation frequencies, which appear in a black hole's response to initial perturbation. The quasinormal modes of black holes do not depend upon the way of the mode's excitation but only on the black hole parameters and are, thereby, ``fingerprints'' of a black hole. It can be shown \cite{Konoplya:2005et} that the response of a wormhole to initial perturbations also is dominated by the quasinormal modes. The quasinormal modes are defined mathematically in the same way for wormholes as they are for the black holes \cite{Konoplya:2005et}. The S-matrix for such problems can be fully described by the wave's reflection/transmission coefficients. We shall show that both the quasinormal modes and reflection coefficients of traversable wormholes are determined by the geometry of a wormhole near its throat. The geometry far from the throat is insignificant for scattering in the same manner as it happens for black holes \cite{Konoplya:2006ar}.

Rotating black holes are known to be spending their rotational energy on amplification of incident waves of perturbation \cite{Starobinsky1}.
This phenomenon occurs also for various rotating compact bodies, such as, for example, conducting cylinders, and is called superradiance \cite{Zeldovich,Starobinsky1,Starobinsky2}.
When considering rotating traversable wormholes, one could probably expect that the same superradiance should take place.
However, we shall show here that rotating axially symmetric traversable wormholes do not allow for the superradiance.

The paper is organized as follows: Sec. II gives some basic information about traversable wormholes. Sec. III is devoted to quasinormal modes
of various traversable wormholes. Sec. IV is about correlation of the scattering coefficients with a wormhole's shape. Sec. V considers scattering around rotating traversable axially symmetric wormholes. Finally, in the Appendix we demonstrate the utility of the 6th order WKB formula for calculations of the graybody factors of black holes, which gives quite accurate ``automatized'' formulas for calculations of the intensity of the Hawking radiation.

\section{Basic features of traversable wormholes}

Static spherically symmetric Lorentzian traversable wormholes of an arbitrary shape can be modeled by a Morris-Thorne ansatz \cite{Morris-Thorne}
\begin{equation}\label{MT}
ds^2 = - e^{2 \Phi (r)} dt^2 + \frac{d r^2}{1 - \frac{b(r)}{r}} + r^2 (d \theta^2 + \sin^2 \theta d \phi^2).
\end{equation}
Here $\Phi(r)$ is the lapse function which determines the red-shift effect and tidal force of the wormhole space-time.
The $\Phi = 0$ wormholes do not produce tidal force.  The shape of a wormhole is completely determined by another function
$b(r)$, called the shape function. When embedding the wormhole metric into the Euclidian space-time with cylindrical coordinates on it,
the equation that describes the embedded surface on the equatorial slice $\theta = \pi/2$ is
\begin{equation}
\frac{dz}{dr} = \pm \left(\frac{r}{b(r)} - 1\right)^{-\frac{1}{2}}.
\end{equation}
The throat of a wormhole is situated at a minimal value of $r$, $r_{min} = b_0$. Thus the coordinate $r$ runs from $r_{min}$ until spatial infinity $r = \infty$, while in the proper radial distance coordinate $d l$, given by the equation
\begin{equation}
\frac{dl}{d r} = \pm \left(1- \frac{b(r)}{r}\right)^{-1/2},
\end{equation}
there are two infinities $l= \pm \infty$ at $r= \infty$.

From the requirement of the absence of singularities, $\Phi(r)$ must be finite everywhere, and the requirement of asymptotic flatness gives $\Phi(r) \rightarrow 0$ as $r \rightarrow \infty $ (or $ l \rightarrow \pm \infty$). The shape function $b(r)$ must be such that $1- b(r)/r > 0$ and $b(r)/r \rightarrow 0$ as $r \rightarrow \infty $ (or $ l \rightarrow \pm \infty$). In the throat $r = b(r)$ and thus $1- b(r)/r$ vanishes. The metric of traversable wormholes does not have a singularity in the throat and the traveler can pass through the wormhole during the finite time. Some thermodynamic properties of traversable wormholes have been recently considered in \cite{Konoplya:2010ak}.

As an illustrative example, we shall consider here two-parametric family of wormholes which are described by the following shape and redshift functions:
\begin{equation}\label{shapefunction}
b(r) = b_{0}^{1-q} r^{q},~q < 1\quad \Phi(r) = \frac{1}{r^p}, ~p > 0 \quad \mbox{or} \quad \Phi(r) = 0
\end{equation}
The shapes $z(r)$ are plotted for the above shape function in Fig. 1. The larger $q$ corresponds to a smaller slope of the curve $z(r)$ respectively horizontal axe $z$, that is a longer spatial extension of the throat regime. For all wormholes $z'(r) = \infty$ \emph{at the throat} $r=b_0$, while for the ultimate case $q \rightarrow 1-0$, $z'(r)$ grows without bound at all points $r$, thus extending the spatial region of the throat dominance.

In principle we could choose other forms of the function $b(r)$ than those we mentioned. This would not change any of the conclusions of this work, because the natural generic form of the wormhole shape (when embedded on the equatorial slice) is some axially symmetric smooth surface with monotonically increasing section when moving away from the throat. This generic form of a wormhole can be represented by the functions (\ref{shapefunction}) or by some other functions satisfying the above mentioned general requirements for traversable wormholes without any qualitative changes of the scattering characteristics.

\begin{figure}
\includegraphics[width=\linewidth]{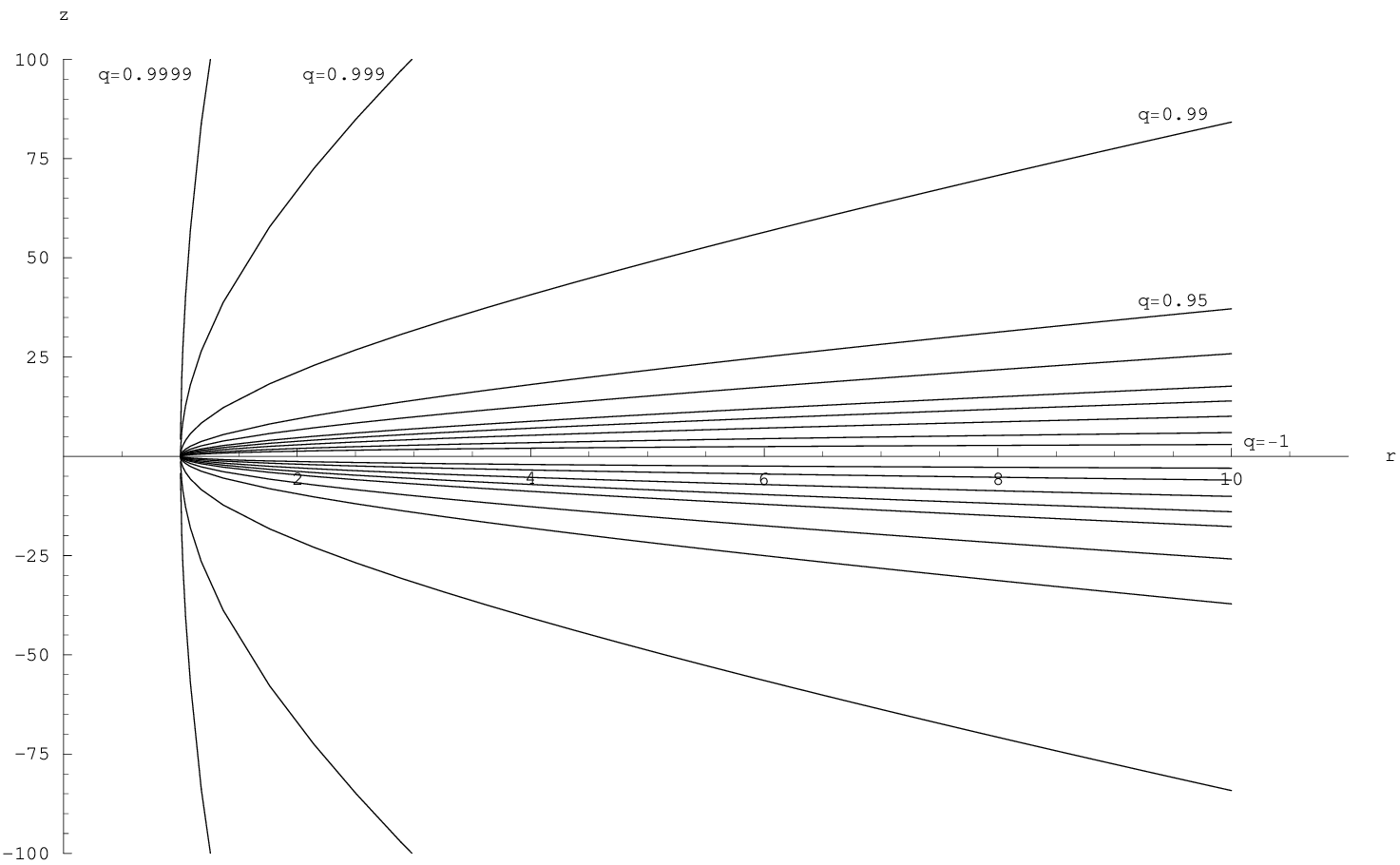}
\caption{Wormhole shapes $z(r)$; $b(r) = b_{0}^{1-q} r^{q}$, $\Phi =0$.}
\end{figure}

\begin{figure}
\includegraphics[width=\linewidth]{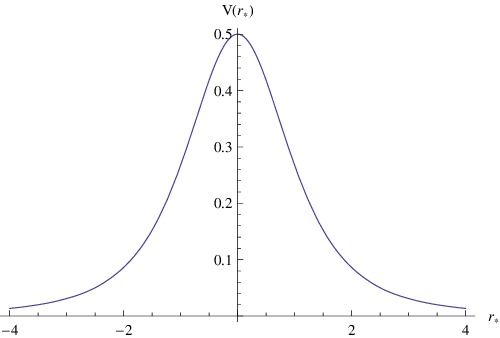}
\caption{The effective potential for the scalar field $\ell=0$ as a function of the tortoise coordinate; $b(r) = b_{0}$, $\Phi =0$.}\label{epl0q0}
\end{figure}

\section{Quasinormal modes}

The wave equations for test scalar $\Phi$ and electromagnetic $A_{\mu}$ fields are
\begin{equation}
(g^{\nu \mu}
\sqrt{-g} \Phi_{,\mu})_{,\nu} = 0,
\end{equation}

\begin{equation}
((A_{\sigma, \alpha} - A_{\alpha, \sigma}) g^{\alpha \mu} g^{\sigma \nu}
\sqrt{-g})_{, \nu} = 0.
\end{equation}
After making use of the metric coefficients (\ref{MT}), the perturbation equations can be reduced to the wavelike form for the scalar and electromagnetic wave functions $\Psi_{s}$ and $\Psi_{el}$ (see for instance \cite{Konoplya:2006ar}):
\begin{equation}\label{wavelike}
\frac{d^{2} \Psi_{i}}{d r_{*}^{2}} + (\omega^{2} - V_{i}(r))\Psi_{i} =
0,\qquad d r_{*}= (A(r) B(r))^{-1/2} dr,
\end{equation}
where the scalar and electromagnetic effective potentials have the form:
\begin{equation}\label{sp}
V_{s} = A(r)\frac{\ell (\ell + 1)}{r^{2}} + \frac{1}{2 r}(A(r) B'(r) +A'(r) B(r)),
\end{equation}
\begin{equation}\label{emp}
V_{el} = A(r) \frac{ \ell (\ell + 1)}{r^{2}}.
\end{equation}
Here $A(r) = e^{2 \Phi(r)}$, $B(r) = 1- b(r)/r$. The effective potentials for the wormholes described by the metric (\ref{MT}), (\ref{shapefunction}) in terms of the tortoise coordinate have the form of the positive definite potential barriers with the peak situated at the throat of a wormhole (see Fig. \ref{epl0q0}). The whole space lies between the two ``infinities'' connecting two universes or distant regions of space. According to \cite{Konoplya:2005et}, the quasinormal modes of wormholes are solutions of the wave equation (\ref{wavelike}) satisfying the following boundary conditions:
\begin{equation}
\Psi \sim e^{ \pm i \omega r_{*}}, \quad r_{*} \rightarrow \pm \infty.
\end{equation}
This means the pure outgoing waves at both infinities or no waves coming from either the left or the right infinity. This is a quite natural condition if one remembers that quasinormal modes are \emph{proper} oscillations of wormholes, i.~e. they represent a response to the perturbation when the ``momentary'' initial perturbation does not affect the propagation of the response at later time. We shall write a quasinormal mode as
\begin{equation}
\omega = \omega_{Re} + i \omega_{Im},
\end{equation}
where $\omega_{Re}$ is the real oscillation frequency and $\omega_{Im}$ is proportional to the decay rate of a given mode.

It is remarkable that the effective potential as a function of the $r$ coordinate (see formulas \ref{emp}, \ref{sp}) from  $b_0$ to infinity has a breaking point at the throat. Though, there is a smooth behavior at the throat respectively the $r_{*}$ coordinate, so that the left and right derivatives of the effective potential in its maximum (throat) are equal. Therefore for finding the quasinormal modes and the $S$-matrix one can use the WKB formula suggested by  Will and Schutz \cite{WKB} and developed to higher orders in \cite{WKBorder}. The same formula was effectively used in a number of works for quasinormal modes of black holes \cite{WKBuse} as well as for some scattering problems \cite{Konoplya:2009hv}.

The 6th order WKB formula reads
\begin{equation}\label{WKBformula}
\frac{\imath Q_{0}}{\sqrt{2 Q_{0}''}} - \sum_{i=2}^{i=6} \Lambda_{i} = N+\frac{1}{2},\qquad N=0,1,2\ldots,
\end{equation}
where $Q = \omega^2 - V$ and the correction terms $\Lambda_{i}$ were obtained in \cite{WKB}, \cite{WKBorder}. Here $Q_{0}^{i}$ means the i-th derivative of $Q$ at its maximum with respect to the tortoise coordinate $r_\star$.

The 6th order WKB quasinormal modes are shown in Figs. 3-6. In Fig. 3 one can see that as $q$ increases from zero to unity, the real part of $\omega$ for the scalar field is monotonically decreasing approaching some constant. For the electromagnetic field $\omega_{Re}$ monotonically increases with $q$ reaching some constant at $q=1$. The damping rate (Figs. 3-6) is monotonically decreasing as $q$ increases approaching asymptotically zero for $q =1$. This is a remarkable result because in the limit of the cylindrical form of geometry near the throat $q=1$ the quasinormal modes reduce to the pure real, nondecaying, quasinormal modes. Such nondecaying pure real modes appear for instance in the spectrum of the massive scalar field of Schwarzschild \cite{Konoplya:2004wg} and Kerr \cite{Konoplya:2006br} black holes and are called \emph{quasiresonances} \cite{quasi-resonance}. Let us note that as we neglected the behavior of the effective potential far from the throat, the approaching of the regime of pure real modes is apparently never exact and in practice we can talk about \emph{long-lived modes} when approaching the limit $b'(r=b_0) =1$.

We can reproduce the above numerical results for higher $\ell$ in the analytical form. Using the first order WKB formula for the general effective potential $V$ with arbitrary $b(r)$ and $\Phi(r)$ and then expanding the result in powers of $1/\ell$ we obtain a relatively concise formula
\JHEPonly{
\begin{equation}
\omega = \frac{e^{\Phi(b_0)}}{b_0} \left(\ell + \frac{1}{2}\right) -\imo\frac{e^{\Phi(b_0)}}{\sqrt{2} b_0} \left(n+\frac{1}{2}\right)
\sqrt{(b'(b_0) -1)(b_0 \Phi'(b_0) -1)}.
\end{equation}
}
\PRDonly{
\begin{eqnarray}
\omega &=& \frac{e^{\Phi(b_0)}}{b_0} \left(\ell + \frac{1}{2}\right) \\\nonumber&&-\imo\frac{e^{\Phi(b_0)}}{\sqrt{2} b_0} \left(n+\frac{1}{2}\right)
\sqrt{(b'(b_0) -1)(b_0 \Phi'(b_0) -1)}.
\end{eqnarray}
}
%$$ K^2 = b_0 (7 + b_0 (\Phi'(b_0) (6 b_0 \Phi`(b_0)-11) + 2b_0 \Phi''(b_0)))$$
%\begin{equation}
%-b_0 (b'(b0) (1- b_0 \Phi'(b_0)) +2(3 b_0 (\Phi'(b_0)(3 b_0\Phi'(b_0) -5) + b_0 \Phi''(b_0))))
%\end{equation}
In the particular case of no-tidal force wormholes ($\Phi =0$), we have
\begin{equation}
\omega = \frac{1}{b_0} \left(\ell + \frac{1}{2}\right) - \imo\frac{1}{\sqrt{2} b_0} \left(n+\frac{1}{2}\right) \sqrt{1 - b^{\prime}(b_0)},
\end{equation}
These formulas are valuable because they usually give good accuracy already for moderately low multipoles $\ell = 2, 3, 4, \ldots$. The above formulas  let us make the two main conclusions: first is that the quasinormal modes are indeed determined by the behavior of the shape function and shiftfunction only near the throat, and second, the limit $b'(r=b_0)=1$ gives pure real nondamping modes, similar to the standing waves of an oscillating string with fixed ends.

Although we considered here only scalar and vector quasinormal modes, a qualitatively similar phenomena could be expected for gravitational perturbations at least for relatively large values of $\ell$ because the centrifugal part of effective potential, which dominates at larger $\ell$, is the same for fields of any spin. Though, as we mentioned earlier, gravitational modes can give surprises in the quasinormal spectrum due to
the exotic character of matter supporting a wormhole.

\begin{figure*}
\centerline{\includegraphics[width=.5\linewidth]{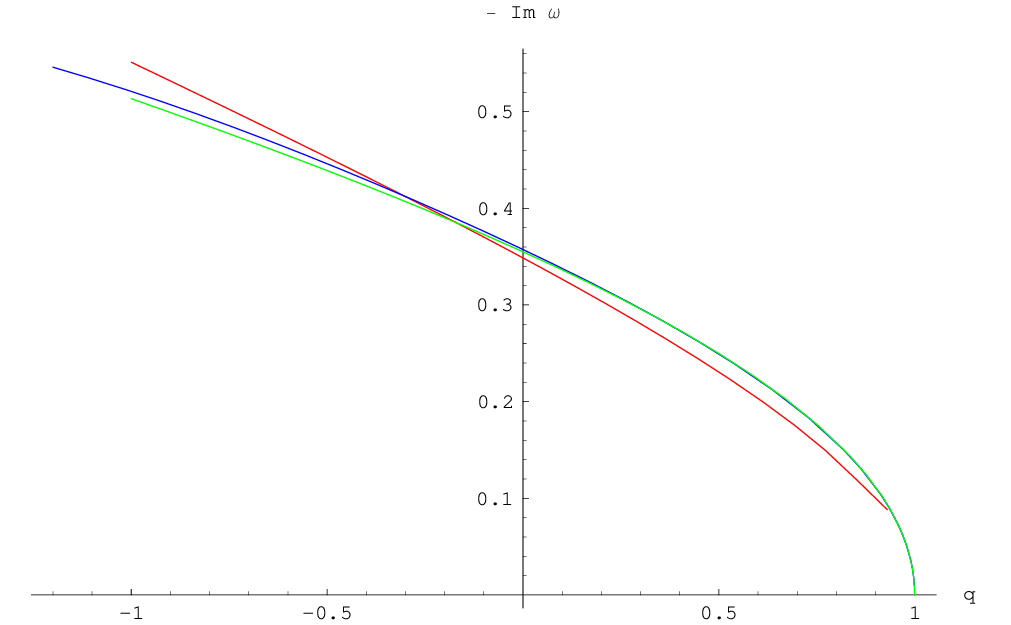}
\includegraphics[width=.5\linewidth]{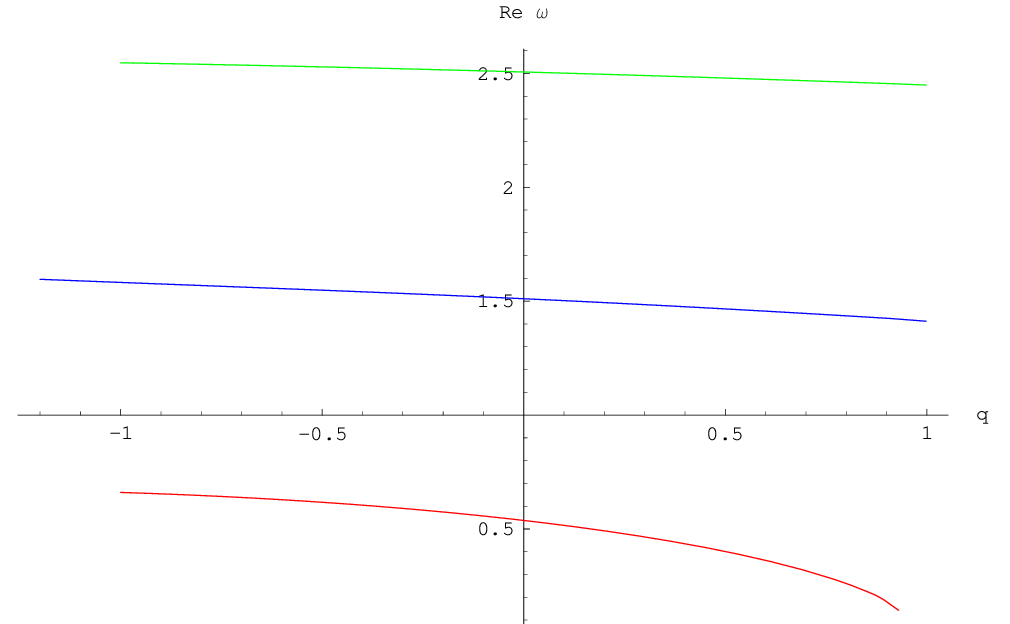}}
\caption{Real (right panel) and imaginary (left panel) parts of quasinormal modes for scalar field for $\ell =0$ (red, bottom), $\ell=1$ (blue), $\ell=2$ (green, top); $b(r) = b_{0}^{1-q} r^{q}$, $\Phi =0$.}
\end{figure*}

\begin{figure*}
\centerline{\includegraphics[width=.5\linewidth]{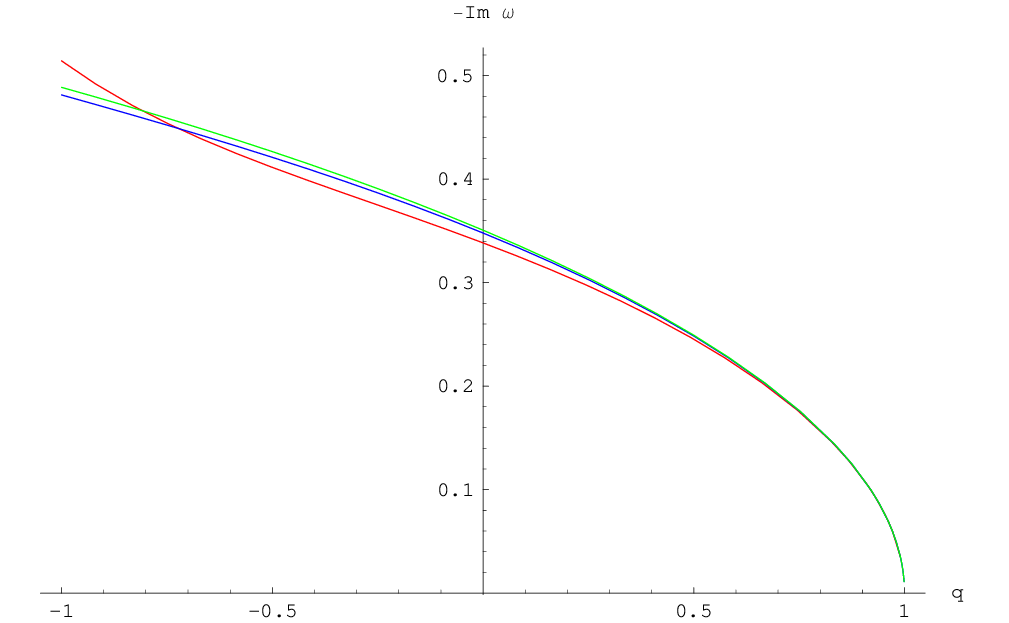}
\includegraphics[width=.5\linewidth]{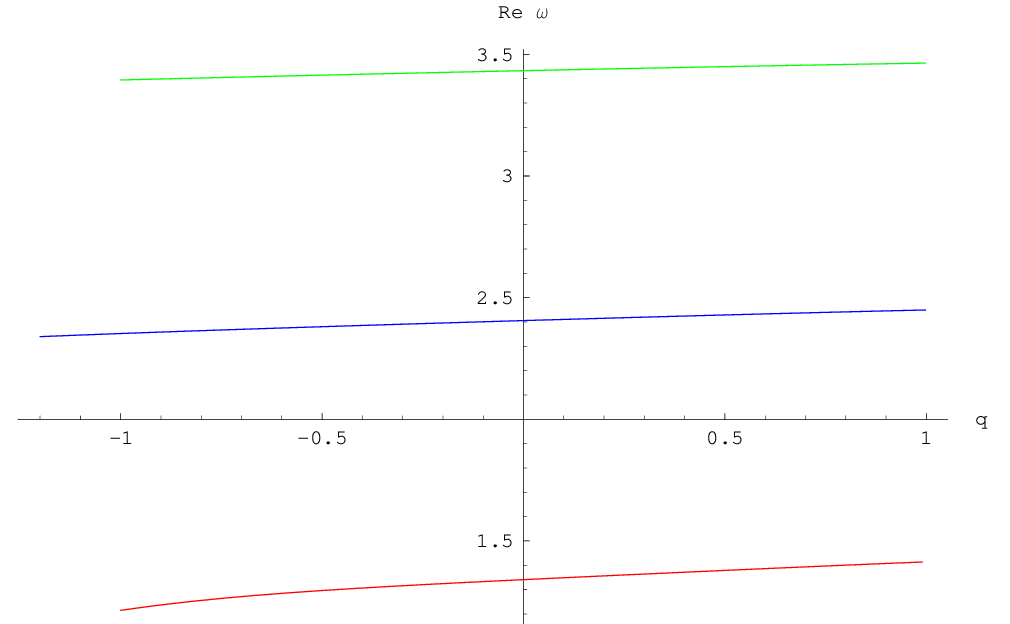}}
\caption{Real (right panel) and imaginary (left panel) parts of quasinormal modes for the Maxwell field for $\ell =1$ (red, bottom), $\ell=2$ (blue), $\ell=3$ (green, top); $b(r) = b_{0}^{1-q} r^{q}$, $\Phi =0$.}
\end{figure*}

\begin{figure*}
\centerline{\includegraphics[width=.5\linewidth]{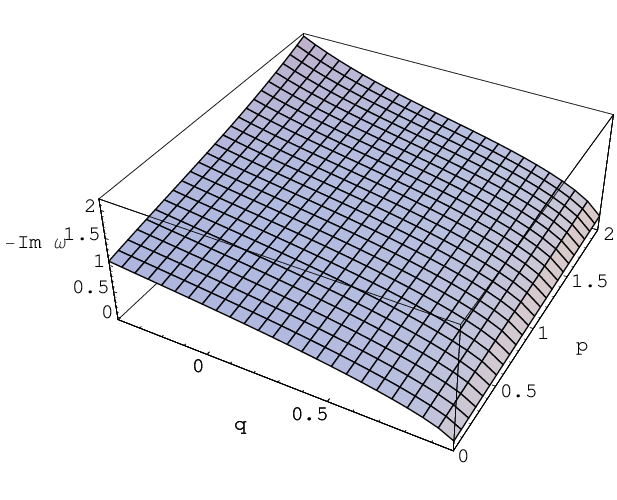}
\includegraphics[width=.5\linewidth]{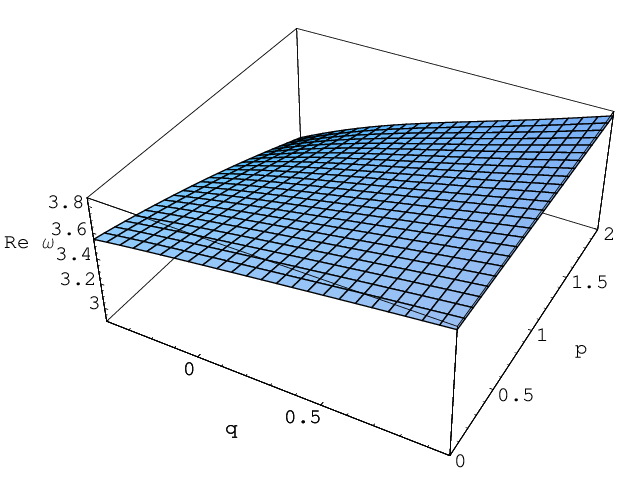}}
\caption{Real (right panel) and imaginary (left panel) parts of quasinormal modes for the Maxwell field for $\ell =1$;
$b(r) = b_{0}^{1-q} r^{q}$, $\Phi =1/r^{p}$.}
\end{figure*}

\begin{figure*}
\centerline{\includegraphics[width=.5\linewidth]{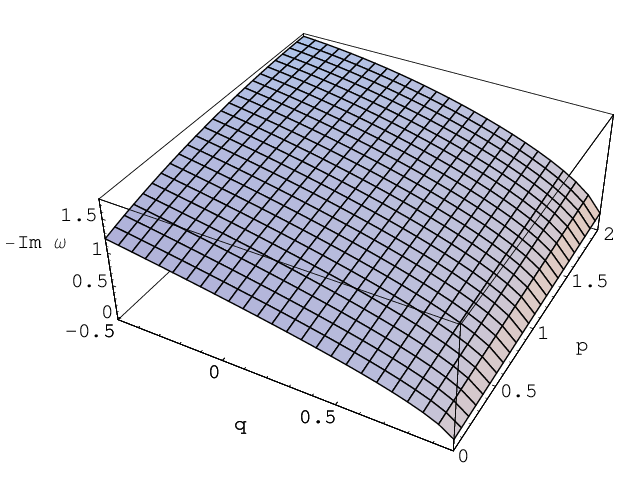}
\includegraphics[width=.5\linewidth]{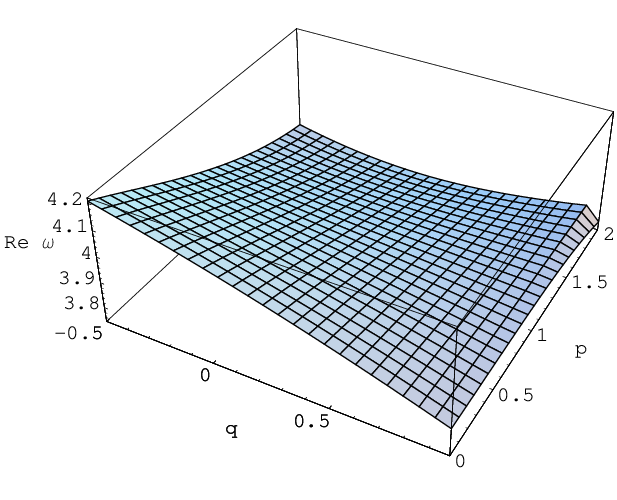}}
\caption{Real (right panel) and imaginary (left panel) parts of quasinormal modes for scalar field for $\ell =1$;
$b(r) = b_{0}^{1-q} r^{q}$, $\Phi =1/r^{p}$}
\end{figure*}

\section{$S$-matrix}

In this section we shall consider the wave equation (\ref{wavelike}) with different boundary conditions, allowing for incoming waves from one of the infinities. This corresponds to the scattering phenomena. The scattering boundary conditions for (\ref{wavelike}) have the following form:
\begin{equation}\label{BC}
\begin{array}{ccll}
    \Psi_i &=& e^{-i\omega r_*} + R e^{i\omega r_*},& r_* \rightarrow +\infty, \\
    \Psi_i &=& T e^{-i\omega r_*},& r_* \rightarrow -\infty, \\
\end{array}%
\end{equation}
where $R$ and $T$ are called the reflection and transmission coefficients.

The above boundary conditions (\ref{BC}) are nothing but the standard scattering boundary conditions for finding the $S$-matrix. The effective potential has the distinctive form of the potential barrier, so that the WKB approach \cite{WKB} can be applied for finding $R$ and $T$. Let us note, that as the wave energy (or frequency) $\omega$ is real, the first order WKB values for $R$ and $T$ will be real \cite{WKB} and
\begin{equation}\label{1}
\left|T\right|^2 + \left|R\right|^2 = 1.
\end{equation}

For $\omega^2 \approx V_{0}$, we shall use the first order beyond the eikonal approximation WKB formula, developed by B.~Schutz and C.~Will (see \cite{WKB}) for scattering around black holes
\begin{equation}\label{WKB1}
R = \left(1 + e^{- 2 i \pi K}\right)^{-1/2},
\quad
\omega^2 \simeq V_{0},
\end{equation}
where $K$ is given in the Appendix.

After the reflection coefficient is calculated we can find the transmission coefficient for the particular multipole number $\ell$
\begin{equation}
\left|{\cal
A}_{\ell}\right|^2=1-\left|R_{\ell}\right|^2=\left|T_{\ell}\right|^2.
\end{equation}

From the Figs. 7 and 8 one can see that in the limit $b'(b_0) =1$, the transmission coefficient approaches asymptotically the theta function, which is dependent on the values of the shift function $\Phi(b_0)$ (see Fig. 9) and the shape function $b_0$ on the throat,
\begin{equation}
\left|T_{\ell}\right|^2 \rightarrow \theta(\omega^2 b_0^2 e^{-2\Phi(b_0)}-\ell(\ell+1)), \quad q \rightarrow 1.
\end{equation}

In the Appendix one can see that the application of the higher order WKB approach allows us to achieve high accuracy in calculations of the transmission coefficients, and, as a result, of the intensity of the Hawking radiation.

\begin{figure*}
\centerline{\includegraphics[width=.5\linewidth]{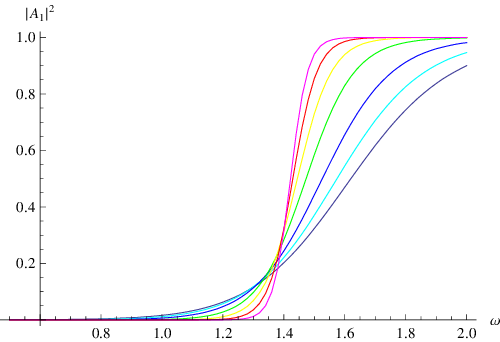}\includegraphics[width=.5\linewidth]{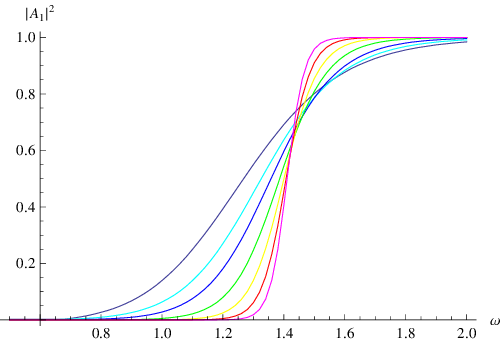}}
\centerline{\includegraphics[width=.5\linewidth]{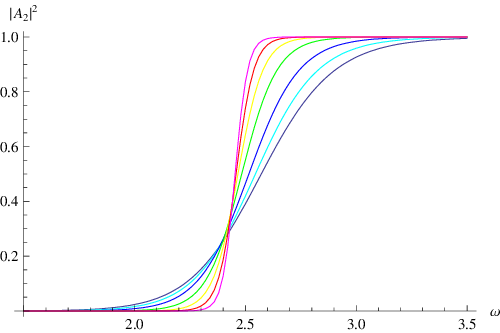}\includegraphics[width=.5\linewidth]{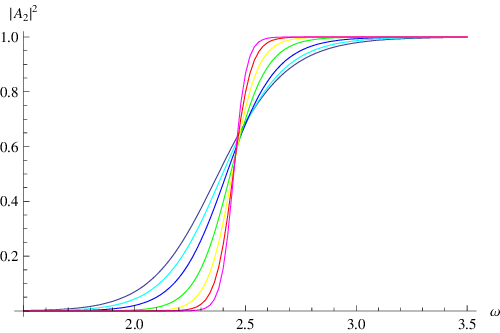}}
\caption{Transmission coefficients for scalar (left panels) and electromagnetic (right panels) fields $\ell =1$ (top panels) and $\ell=2$ (bottom panels), $\Phi =0$, $b(r) = b_{0}^{1-q} r^{q}$, $q=-1$ (violet), $q=-1/2$ (cyan), $q=0$ (blue), $q=1/2$ (green), $q=3/4$ (yellow), $q=7/8$ (red), $q=15/16$ (magenta). As $q\rightarrow1$ the transmission coefficient approaches the theta function $\theta(\omega^2 b_0^2-\ell(\ell+1))$.}
\end{figure*}

\begin{figure*}
\centerline{\includegraphics[width=.5\linewidth]{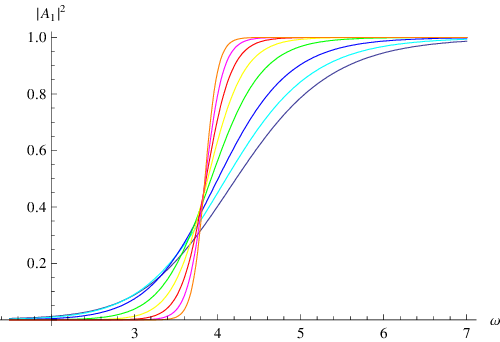}\includegraphics[width=.5\linewidth]{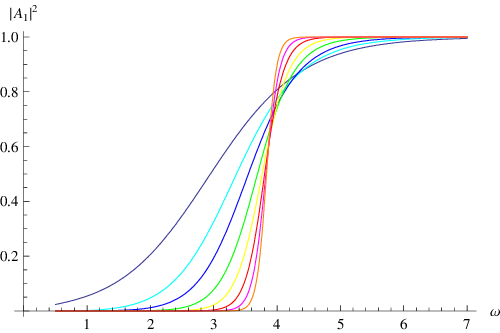}}
\centerline{\includegraphics[width=.5\linewidth]{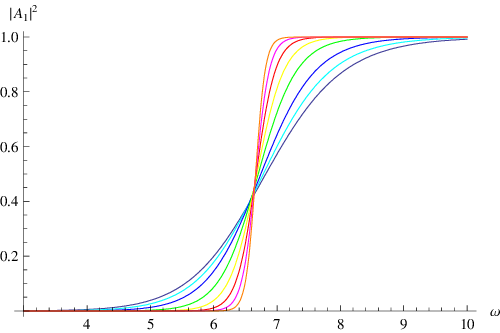}\includegraphics[width=.5\linewidth]{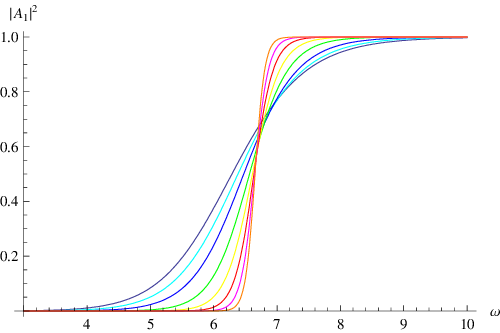}}
\caption{Transmission coefficient for scalar (left panels) and electromagnetic (right panels) fields $\ell =1$ (top panels) and $\ell=2$ (bottom panels), $\Phi =\frac{b_0}{r}$, $b(r) = b_{0}^{1-q} r^{q}$, $q=-1$ (violet), $q=-1/2$ (cyan), $q=0$ (blue), $q=1/2$ (green), $q=3/4$ (yellow), $q=7/8$ (red), $q=15/16$ (magenta), $q=31/32$ (orange). As $q\rightarrow1$ the transmission coefficient approaches the theta function $\theta(\omega^2 b_0^2 e^{-2\Phi(b_0)}-\ell(\ell+1))$.}
\end{figure*}

\begin{figure*}
\centerline{\includegraphics[width=.50\linewidth]{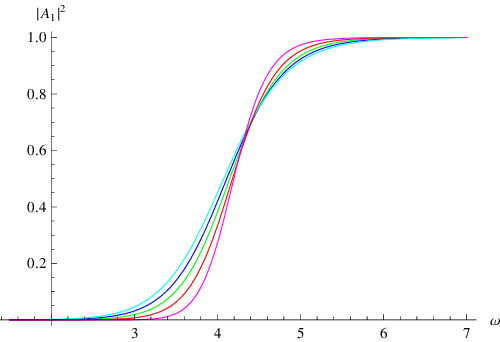}\includegraphics[width=.50\linewidth]{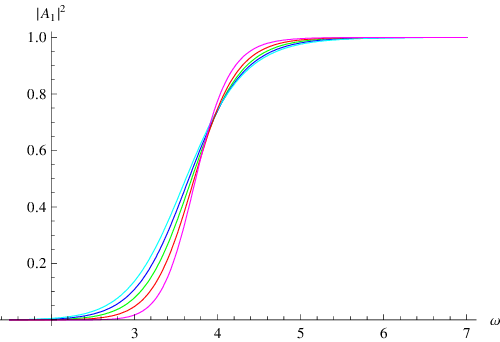}}
\caption{Transmission coefficient for scalar (left) and electromagnetic (right) fields $\ell =1$, $b(r) = b_{0}$, $\Phi =
\left(\frac{b_0}{r}\right)^p$: $p=-1/2$ (magenta), $p=-1/4$ (red),
$p=0$ (green), $p=1/4$ (blue), $p=1/2$ (cyan). The more sloping
line corresponds to the lower value of $p$.}
\end{figure*}

\section{Rotating axisymmetric traversable wormholes}

The asymptotic forms of the solutions near the black hole and at the spatial infinity are
\begin{equation}\label{infty-asym}
\Psi(r_{*}) = e^{-i \omega r_{*}} + R e^{+ i \omega r_{*}}, \quad r \rightarrow \infty,
\end{equation}
\begin{equation}\label{horizon-asym}
\Psi(r_{*}) = T e^{- i (\omega - m \Omega_{h}) r_{*}}, \quad r \rightarrow r_{+}.
\end{equation}
Here $R$ is called the amplitude of the reflected wave or the reflection coefficient, and $T$ is the transmission coefficient. If $|R| > 1$, that is
\begin{equation}\label{superrad-cond}
\frac{m \Omega_h}{\omega} > 1,
\end{equation}
then, the reflected wave has larger amplitude than the incident one. This amplification of the incident wave is called the \emph{superradiance} and was first predicted by Zel'dovich \cite{Zeldovich}. The superradiance effect for Kerr black holes was first calculated by Starobinsky \cite{Starobinsky1,Starobinsky2}. The process of superradiant amplification occurs due to the extraction of rotational energy of a black hole, and therefore, it happens only for modes with positive values of azimuthal number $m$, that corresponds to ``corotation'' with a black hole. Some aspects of super-radiance of four dimensional black holes of astrophysical interest were considered in \cite{superradianceBH}.

Let us now see if this superradiant amplification is possible for traversable wormholes. The line element of the stationary, axially symmetric traversable wormhole can be written as \cite{Teo:1998dp}
\begin{equation}
ds^2=-e^{2\Phi}dt^2 +\frac{dr^2}{1 -b/r}+r^2K^2(d\theta^2+\sin^2\theta(d\phi-{\bar \omega} dt)^2),
\end{equation}
where $\Phi$, $b$, $K$, and ${\bar \omega}$, being functions of $r$ and $\theta$, are chosen in such a way that they are regular on the symmetry axis $\theta = 0, \pi$.

When $\Phi$, $b$, and ${\bar \omega}$ are functions of $r$ and $K=K(\theta)$ we can separate the variables in the equation of motion for the test scalar field \cite{Kim:2004ph}
\begin{equation}
\frac{1}{\sqrt{-g}}\partial_\mu (
g^{\mu\nu}\sqrt{-g}\partial_\nu \phi) = 0.
\end{equation}
Using the ansatz
$$\phi = \frac{\Psi(r)}{r}S(\theta)e^{im\phi}e^{i\omega t}$$
we find the wavelike equation
\begin{equation}
\frac{d^2\Psi}{dr_*^2}+(\omega+m{\bar \omega})^2\Psi-\left(\frac{e^{2\Phi}\lambda_{\ell,m}}{r^2}+\frac{1}{r}\frac{de^{\Phi}\sqrt{1 -\frac{b}{r}}}{dr_*}\right)\Psi = 0,
\end{equation}
where $\lambda_{\ell,m}$ is the angular separation constant, which depends on the azimuthal number $m$ and the multipole number $\ell\geq|m|$. In the particular case, if $K=1$, $S(\theta)$ are spheroidal functions, and $\lambda_{\ell,m}=\ell(\ell+1)$.

The tortoise coordinate is given by
$$dr_*=\frac{dr}{e^{\Phi}\sqrt{1 -\frac{b}{r}}}.$$

The metric describes two identical, asymptotically flat regions joined together at the throat $r = b > 0$. Therefore, the boundary conditions for the radial part of the function at the infinities are symmetric
$$\Psi(r)\propto e^{\pm\imo\Omega(\omega,m)r}, \qquad r\rightarrow\infty,$$
where $\Omega(\omega,m)=\omega+m{\bar \omega}(r=\infty)$.

The tortoise coordinate $r_*$ maps the two regions $r>b$ onto $(-\infty,\infty)$ and for the scattering problem we have the following boundary conditions:
\begin{eqnarray}\nonumber
\Psi&=&e^{-\imo\Omega(\omega,m)r_*}+R e^{\imo\Omega(\omega,m)r_*}, \qquad r_*\rightarrow\infty,\\\nonumber
\Psi&=&Te^{-\imo\Omega(\omega,m)r_*}, \qquad\qquad\qquad r_*\rightarrow-\infty.
\end{eqnarray}

By comparing the Wronskian of the two linearly independent solutions of the wavelike equation $\Psi$ and $\Psi^*$ at the boundaries, we find that
$$|T|^2+|R|^2=1$$
for any possible effective potential.

This relation does not depend on the particular form of the effective potential and the function ${\bar \omega}$ and can be also found for other test fields and gravitational perturbations. Thus, we conclude that there is no superradiance for axially symmetric traversable wormholes. Because of the symmetry of the regions at both sides of the wormhole, the energy, extracted from the wormhole when a wave falls inside it, is absorbed by the wormhole on the other side.

\section{Discussions}

We have considered the propagation of test scalar and Maxwell fields in the background of the traversable spherically symmetric and axially symmetric wormholes of a generic form. The majority of our conclusions are independent of the concrete shape of a wormhole. We have found that when the shape function $b(r)$ approaches the regime $b'(r) =1$ at the wormhole's throat $b=b_0$, the quasinormal modes asymptotically approache pure real nondecaying modes, meaning the appearance of long-lived modes. The latter might give much better observational opportunities. The reflection coefficient approaches the step function which depends on the values of the shape and redshift function at the throat. We have proved that independently on their particular forms, traversable axially symmetric wormholes do not allow for superradiance. We were limited here by wormholes whose shape is symmetric with respect to their center (throat). Spherically (but not axially) symmetric wormholes are also necessarily symmetric with respect to the throat.

\section*{Acknowledgments}
At the earlier stage this work was supported by the {\it Japan Society for the Promotion of Science (JSPS)}, Japan and in the final stage by \emph{the Alexander von Humboldt Foundation}, Germany.
A. Z. was supported by \emph{Funda\c{c}\~ao de Amparo \`a Pesquisa do Estado de S\~ao Paulo (FAPESP)}, Brazil.

\begin{figure*}
\centerline{\includegraphics[width=.5\linewidth]{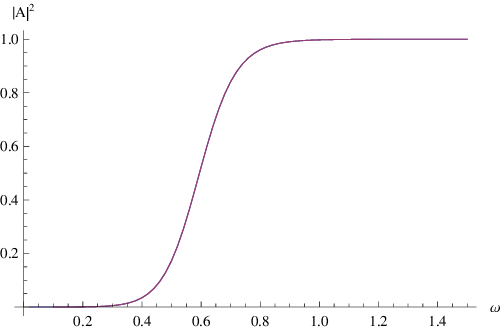}\includegraphics[width=.5\linewidth]{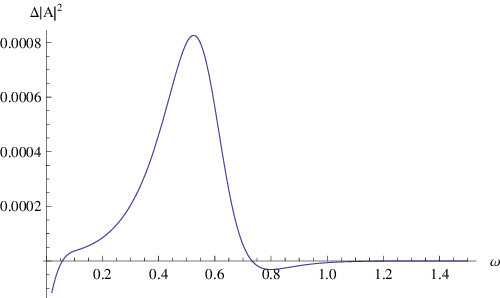}}
\caption{On the left figure we show the graybody factors for the scalar field $\ell=1$ in the Schwarzschild background found with the help of the WKB formula (red line) and the accurate values found by the shooting method (blue line). The right figure shows the difference between the WKB formula and the accurate results.}\label{A1WKB}
\end{figure*}

\begin{figure}
\includegraphics[width=\linewidth]{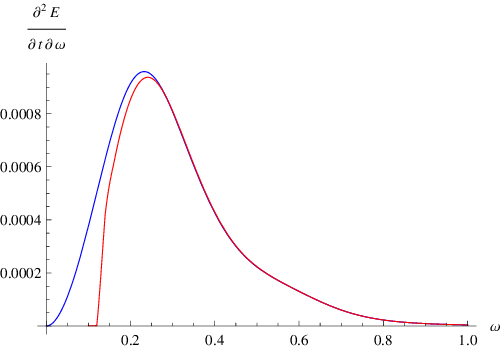}
\caption{Energy emission rate of the Schwarzschild black hole for the scalar field found with the help of the WKB formula (red, bottom) and the accurate values obtained by the shooting method (blue, top).}\label{EERWKB}
\end{figure}

\section*{Appendix: Calculation of the reflection coefficients for generic black and wormholes by the 6th order WKB formula}

In order to estimate accuracy of the WKB formula for the scattering problem, we shall consider the well-known Schwarzschild metric
\begin{equation}
ds^2=-f(r)dt^2 +\frac{dr^2}{f(r)}+r^2(d\theta^2+\sin^2\theta d\phi^2), ~ f(r)=1-\frac{1}{r}.
\end{equation}
The scalar field equation,
$$\Box\phi=0,$$
can be reduced to the wavelike equation (\ref{wavelike}) with the effective potential
$$V(r)=f(r)\left(\frac{\ell(\ell+1)}{r^2}+\frac{f'(r)}{r}\right).$$

The reflection coefficient, given by the WKB formula, is
\begin{equation}\label{moderate-omega-wkb}
R = (1 + e^{- 2 i \pi K})^{-\frac{1}{2}},
\end{equation}
where
\begin{equation}
K = i \frac{(\omega^2 - V_{0})}{\sqrt{-2 V_{0}^{\prime \prime}}} + \sum_{i=2}^{i=6} \Lambda_{i}.
\end{equation}
Here $V_0$ is the maximum of the effective potential, $V_{0}^{\prime \prime}$ is the second derivative of the
effective potential in its maximum with respect to the tortoise coordinate, and $\Lambda_i$  are higher order WKB corrections which depend on up to $2i$th order derivatives of the effective potential at its maximum \cite{WKBorder}.

We compare the graybody factors found with the help of the WKB formula with the graybody factor calculated accurately by fitting the numerical solution of the wavelike equation (see Fig. \ref{A1WKB}). In Fig. \ref{EERWKB} we show the energy emission rate for the scalar field. There one can see that the accuracy of the 6th order WKB formula is exceptionally good for not very small values of $\omega$ giving us at hand an automatic and powerful method for estimation of the Hawking radiation effect for various black holes.

The automatized WKB formula for the quasinormal modes and reflection coefficient in the Mathematica\textregistered{} Notebook format and examples of using the WKB formula for Schwarzschild black hole can be downloaded from \url{https://goo.gl/nykYGL}. This procedure (without any modifications) can be easily applied for calculations of characteristics of the Hawking radiation of various black holes. That is much easier than the shooting method which requires the analysis of asymptotical behavior of the wave equation for each black hole.

One should note that for small $\omega$ the accuracy of the WKB formula is small because of the large distance between the turning points in that case. For higher accuracy for small $\omega$ is achieved by the other formula,
\begin{equation}\label{small-omega-wkb}
T=\exp\left(-\int dr_*\sqrt{V(r_*)-\omega^2}\right),
\end{equation}
where the integration is performed between the two turning points $V(r_*)=\omega^2$. Though the sixth-order WKB formula provides an excellent accuracy for the gravitational perturbations, for which $\ell\geq2$, even for small $\omega$. The matching of both WKB formulas, (\ref{small-omega-wkb}) for small $\omega$ and (\ref{moderate-omega-wkb}) for moderate $\omega$, provides usually very good estimation of the Hawking effect.

\end{document}